\RequirePackage{fix-cm}
\documentclass[smallextended]{svjour3}
\smartqed  
\usepackage[table,xcdraw]{xcolor}
\usepackage{mathptmx}
\usepackage[utf8]{inputenc}
\usepackage{soul,xcolor}
\usepackage{graphicx}
\usepackage{caption}
\usepackage{nicefrac}
\usepackage{amsfonts}
\usepackage[T1]{fontenc}
\usepackage{xspace}
\usepackage{flushend}
\usepackage{ifthen}
\usepackage{color}
\usepackage{url}
\usepackage{multirow}
\usepackage{listliketab}
\usepackage{amssymb}
\usepackage[linesnumbered,lined,boxed,commentsnumbered]{algorithm2e}
\usepackage{algorithmic}
\usepackage{booktabs}
\usepackage{comment}
\usepackage{amsmath}
\usepackage{bbding}
\usepackage{listings}
\usepackage{float}
\usepackage[misc]{ifsym}
\usepackage{tabularx}
\usepackage{rotating}
\usepackage{natbib}
\usepackage{subfigure}
\usepackage{ulem}
\usepackage{censor}
\usepackage{subcaption}
\usepackage{pgfplots}
\pgfplotsset{compat=1.14}
\usepackage{pgfplotstable}
\usepackage{graphicx}% biography <<<<<<<<<<
\usepackage{wrapfig}% biography <<<<<<<<<<  
\PassOptionsToPackage{hyphens}{url}
\usepackage[hidelinks]{hyperref}
\usepackage{tablefootnote}
\usepackage[table]{xcolor}
\usepackage{graphicx, subfigure}
\usepackage{caption}
\usepackage{makecell}
\usepackage{soul}
\usepackage{listings}
\usepackage{multirow}
\usepackage{pifont}
\usepackage{xspace}
\usepackage{epigraph} 
\usepackage[flushleft]{threeparttable}
\usepackage{url}
\definecolor{backcolour}{rgb}{0.95,0.95,0.92}
\usepackage{algorithm2e}
\UseRawInputEncoding
\usepackage[colorinlistoftodos]{todonotes}
\usepackage{listings}
\usepackage{comment}
\usepackage{fontawesome}

\definecolor{celestialblue}{rgb}{0.29, 0.59, 0.82}
\definecolor{awesome}{rgb}{0.0, 0.2, 0.6}
\definecolor{coolblack}{rgb}{0.0, 0.18, 0.39}
\definecolor{maroon}{cmyk}{0, 0.87, 0.68, 0.32}
\definecolor{halfgray}{gray}{0.55}
\definecolor{ipython_frame}{RGB}{207, 207, 207}
\definecolor{ipython_bg}{RGB}{247, 247, 247}
\definecolor{ipython_red}{RGB}{186, 33, 33}
\definecolor{ipython_green}{RGB}{0, 128, 0}
\definecolor{ipython_cyan}{RGB}{64, 128, 128}
\definecolor{ipython_purple}{RGB}{170, 34, 255}

\definecolor{Fork}{HTML}{99CCFF}
\definecolor{Non-Fork}{HTML}{FFCCCB}

\newcommand\revise[1]{{\color{black}#1}}

\definecolor{chestnut}{rgb}{0.8, 0.36, 0.36}
\definecolor{red}{rgb}{1, 0.25, 0.098}
\definecolor{blue}{rgb}{0.25, 0.44, 1}
\definecolor{purple}{rgb}{0.631, 0.475, 0.949}
\definecolor{pink}{rgb}{0.969, 0.341, 0.549}
\definecolor{navy}{rgb}{0.09, 0.10, 0.19}
\definecolor{light}{rgb}{0.71, 0.73, 0.77}

\usepackage{tikz}

\colorlet{punct}{red!60!black}
\definecolor{background}{HTML}{EEEEEE}
\definecolor{delim}{RGB}{20,105,176}
\colorlet{numb}{magenta!60!black}

\lstdefinelanguage{json}{
    basicstyle=\small,
    numbers=left,
    numberstyle=\scriptsize,
    stepnumber=1,
    numbersep=8pt,
    showstringspaces=false,
    breaklines=true,
    frame=lines,
    backgroundcolor=\color{background},
    literate=
     *{0}{{{\color{numb}0}}}{1}
      {1}{{{\color{numb}1}}}{1}
      {2}{{{\color{numb}2}}}{1}
      {3}{{{\color{numb}3}}}{1}
      {4}{{{\color{numb}4}}}{1}
      {5}{{{\color{numb}5}}}{1}
      {6}{{{\color{numb}6}}}{1}
      {7}{{{\color{numb}7}}}{1}
      {8}{{{\color{numb}8}}}{1}
      {9}{{{\color{numb}9}}}{1}
      {:}{{{\color{punct}{:}}}}{1}
      {,}{{{\color{punct}{,}}}}{1}
      {\{}{{{\color{delim}{\{}}}}{1}
      {\}}{{{\color{delim}{\}}}}}{1}
      {[}{{{\color{delim}{[}}}}{1}
      {]}{{{\color{delim}{]}}}}{1},
}

\usepackage[strict]{changepage}
  \definecolor{ABlue}{HTML}{127bca}
 \definecolor{LHScolor}{HTML}{555555}
\usepackage{framed}

\newcommand{\droptextshadow}[2]{%
    \tikz[baseline,outer sep=0pt, inner sep=0pt]{
    \node[#1!40!black] at (0,-0.1ex) {#2};
    \node[white] at (0,0) {#2};
}%
}

\newcommand{\DOIbox}[1]{
\tcbsidebyside[
        bicolor,
        sidebyside,
        sidebyside adapt=both,
        sidebyside gap=5pt,
        top=0pt,left=0pt,right=0pt,bottom=0pt,
        boxrule=0pt,rounded corners,
        interior style={top color=LHScolor,bottom color=LHScolor!60!black},
        segmentation style={top color=ABlue,bottom color=ABlue!60!black},
]{%
\droptextshadow{LHScolor}{DOI}% <-- Drop shadow + text for DOI 
}{%
\droptextshadow{ABlue}{\href{http://dx.doi.org/#1}{#1}}% <-- Drop shadow + text for reference number + hyperref
}%
}

\definecolor{formalshade}{rgb}{1.0,1.0,1.0}
\definecolor{side}{rgb}{0.0,0.2,0.6}

\newenvironment{formal}{%
  \MakeFramed{\advance\hsize-\width\FrameRestore}%
  \noindent\hspace{-4.55pt}% disable indenting first paragraph
  \begin{adjustwidth}{}{7pt}%
  \vspace{2pt}\vspace{2pt}%
}
{%
  \vspace{2pt}\end{adjustwidth}\endMakeFramed%
}
\definecolor{gray(x11gray)}{rgb}{0.75, 0.75, 0.75}

\lstdefinelanguage{python}{
    morekeywords={access,and,break,class,continue,def,del,elif,else,except,exec,finally,for,from,global,if,import,in,is,lambda,not,or,pass,print,raise,return,try,while},
    morekeywords=[2]{abs,all,any,basestring,bin,bool,bytearray,callable,chr,classmethod,cmp,compile,complex,delattr,dict,dir,divmod,enumerate,eval,execfile,file,filter,float,format,frozenset,getattr,globals,hasattr,hash,help,hex,id,input,int,isinstance,issubclass,iter,len,list,locals,long,map,max,memoryview,min,next,object,oct,open,ord,pow,property,range,raw_input,reduce,reload,repr,reversed,round,set,setattr,slice,sorted,staticmethod,str,sum,super,tuple,type,unichr,unicode,vars,xrange,zip,apply,buffer,coerce,intern},
    sensitive=true,
    morecomment=[l]\#,
    morestring=[b]',
    morestring=[b]",
    morestring=[s]{'''}{'''},
    morestring=[s]{"""}{"""},
    morestring=[s]{r'}{'},
    morestring=[s]{r"}{"},
    morestring=[s]{r'''}{'''},
    morestring=[s]{r"""}{"""},
    morestring=[s]{u'}{'},
    morestring=[s]{u"}{"},
    morestring=[s]{u'''}{'''},
    morestring=[s]{u"""}{"""},
    literate=
    {á}{{\'a}}1 {é}{{\'e}}1 {í}{{\'i}}1 {ó}{{\'o}}1 {ú}{{\'u}}1
    {Á}{{\'A}}1 {É}{{\'E}}1 {Í}{{\'I}}1 {Ó}{{\'O}}1 {Ú}{{\'U}}1
    {à}{{\`a}}1 {è}{{\`e}}1 {ì}{{\`i}}1 {ò}{{\`o}}1 {ù}{{\`u}}1
    {À}{{\`A}}1 {È}{{\'E}}1 {Ì}{{\`I}}1 {Ò}{{\`O}}1 {Ù}{{\`U}}1
    {ä}{{\"a}}1 {ë}{{\"e}}1 {ï}{{\"i}}1 {ö}{{\"o}}1 {ü}{{\"u}}1
    {Ä}{{\"A}}1 {Ë}{{\"E}}1 {Ï}{{\"I}}1 {Ö}{{\"O}}1 {Ü}{{\"U}}1
    {â}{{\^a}}1 {ê}{{\^e}}1 {î}{{\^i}}1 {ô}{{\^o}}1 {û}{{\^u}}1
    {Â}{{\^A}}1 {Ê}{{\^E}}1 {Î}{{\^I}}1 {Ô}{{\^O}}1 {Û}{{\^U}}1
    {œ}{{\oe}}1 {Œ}{{\OE}}1 {æ}{{\ae}}1 {Æ}{{\AE}}1 {ß}{{\ss}}1
    {ç}{{\c c}}1 {Ç}{{\c C}}1 {ø}{{\o}}1 {å}{{\r a}}1 {Å}{{\r A}}1
    {€}{{\EUR}}1 {£}{{\pounds}}1
    {^}{{{\color{ipython_purple}\^{}}}}1
    {=}{{{\color{ipython_purple}=}}}1
    {+}{{{\color{ipython_purple}+}}}1
    {*}{{{\color{ipython_purple}$^\ast$}}}1
    {/}{{{\color{ipython_purple}/}}}1
    {+=}{{{+=}}}1
    {-=}{{{-=}}}1
    {*=}{{{$^\ast$=}}}1
    {/=}{{{/=}}}1,
    literate=
    *{-}{{{\color{ipython_purple}-}}}1
     {?}{{{\color{ipython_purple}?}}}1,
    identifierstyle=\color{black}\ttfamily,
    commentstyle=\color{ipython_cyan}\ttfamily,
    stringstyle=\color{ipython_red}\ttfamily,
    keepspaces=true,
    showspaces=false,
    showstringspaces=false,
    rulecolor=\color{ipython_frame},
    numberstyle=\tiny\color{halfgray},
    backgroundcolor=\color{ipython_bg},
    basicstyle=\scriptsize,
    keywordstyle=\color{ipython_green}\ttfamily,
}

\usepackage[skins,breakable]{tcolorbox}

\newcommand{\RqOne}{\textbf{RQ1:} \textbf{\ul{To what extent do developers engage in protestware discussions?}} \xspace}

\newcommand{\RqTwo}{\textbf{RQ2:} \textbf{\ul{How can we characterize the contents of discussion regarding protestware between developers?}} \xspace} 

\newcommand{\RqTwoA}{\textbf{RQ2a:} \textbf{\ul{What are the different stances and communicative styles that developers take when discussing protestware?}} \xspace} 
\newcommand{\RqTwoB}{\textbf{RQ2b:} \textbf{\ul{What different themes emerge from the discussions of protestware?}} \xspace} 
\newcommand{\RqTwoC}{\textbf{RQ2c:} \textbf{\ul{What different kinds of mitigation strategies do developers provide for protestware?}} \xspace} 

\newcommand{\RqThree}{\textbf{RQ3:} \textbf{\ul{To what extent do developers abandon protestware in their software?}} \xspace} 

\newcommand{\caseOne}{{colors.js}}
\newcommand{\caseTwo}{{es5-ext}}
\newcommand{\caseThree}{{ua-parser}}
\newcommand{\caseFour}{{log4j}}

\definecolor{Large}{HTML}{696969}
\definecolor{Negligible}{HTML}{D3D3D3}
\definecolor{Medium}{HTML}{808080}
\definecolor{Small}{HTML}{A9A9A9}
\begin{document}

\title{Developer Reactions to Protestware in Open Source Software: The cases of \texttt{color.js} and \texttt{es5.ext}}

\author{Youmei Fan \Letter \and Dong Wang \and Supatsara Wattanakriengkrai  \and Hathaichanok Damrongsiri \and Christoph Treude \and Hideaki Hata \and Raula Gaikovina Kula
}

\institute{
    \Letter~Corresponding author - Youmei Fan\\
    Youmei Fan, Raula Gaikovina Kula, Supatsara Wattanakriengkrai, Hathaichanok Damrongsiri
    \at Nara Institute of Science and Technology, Japan\\
    \email{\{fan.youmei.fs2, raula-k, wattanakri.supatsara.ws3, damrongsiri.hathaichanok.db5\}@is.naist.jp}
    \and
    Dong Wang \at
    College of Intelligence and Computing, Tianjin University, China\\
    \email{dong\_w@tju.edu.cn}
    \and
    Christoph Treude \at
    Singapore management university, Singapore\\
    \email{ctreude@smu.edu.sg}
    \and
    Hideaki Hata \at
    Shinshu University, Japan\\
    \email{hata@shinshu-u.ac.jp}
}

\date{Received: date / Accepted: date}
\setstcolor{red}
\maketitle

\begin{abstract}
There is growing concern about maintainers self-sabotaging their work in order to take political or economic stances, a practice referred to as ``protestware''.
Our objective is to understand the discourse around discussions on such an attack, how it is received by the community, and whether developers respond to the attack in a timely manner. We study two notable protestware cases i.e., \caseOne~and \caseTwo.
Results indicate that \revise{protestware discussions are spread more quickly on the GitHub platform, while security vulnerabilities are faster on social media.}
By establishing a taxonomy of protestware discussions, we identify posts that express stances and provide technical mitigation instructions. We applied a thematic analysis \revise{to 684 protestware related posts} to identify five major themes during the discussions: {i. disseminate and response}, {ii. stance}, {iii. reputation}, {iv. communicative styles}, {v. rights and ethics}.
This work sheds light on the nuanced landscape of protestware discussions, offering insights for both researchers and developers into maintaining a healthy balance between the political or social actions of developers and the collective well-being of the open-source community.
\end{abstract}

\section{Introduction}
\label{sec:introduction}

\begin{formal}
    \color{coolblack}{\textit{``Anyone who experienced actual significant disruption from this (protestware) brought it on themselves with their bad dev practices. No one forced anyone to install the latest version without actually verifying it at all. Didnt corrupt the version history so nope, people just letting their entitlement and lack of understanding of licenses show.'' }} \faGithub
\end{formal}

\begin{formal}
    \color{coolblack}{\textit{``In dev-land, we don't stand on the shoulders of giants. We keep our life-rafts afloat by sticky-taping together skerricks of code that hopefully has more buoyancy than ballast. And sometime it just takes one person to take the whole ship down.'' }} \faGithub
\end{formal}

Open source software development has emerged as an unexpected platform for expressing social and political protests. In recent years, we have witnessed the emergence of ``protestware'', defined as when developers deliberately express dissent and draw attention to issues they consider important~\citep{kula2022war}. 
Protestware takes on many forms. For example, the maintainer of `node-ipc' used malicious code to target host machines with IP addresses in Russia or Belarus in response to the War in Ukraine~\citep{massacci2022free}. 
GitHub declared this a critical vulnerability, known as CVE-2022-23812.\footnote{https://nvd.nist.gov/vuln/detail/cve-2022-23812}
This act underscores the political motivations that can inspire the creation of protestware. Other instances of protestware stem from industry concerns, such as the perceived exploitation of open-source labor by large corporations. For instance, the developer of the `faker' library intentionally introduced an infinite loop, disrupting thousands of projects~\citep{bellovin2022open}. Today, `faker' is a community-controlled project, maintained by a team of developers from various companies~\citep{fakerjsUpdateFrom}.

Such protests can result in widespread consequences, particularly as modern software systems often heavily rely on third-party libraries, rendering them potential points of vulnerability~\citep{zahan2022weak}. To illustrate the scale of this interconnectedness, libraries listed in the popular NPM registry, which hosts over a million libraries, each depend on an average of five to six other libraries within the same ecosystem~\citep{chinthanet2021makes}.
Beyond the technical issues, the rise of protestware threatens the trust that is foundational to modern software ecosystems~\citep{ghofrani2022trust}, creating disruptions within the software development community. Developers using a protesting library must decide whether to continue using it or take on the potentially complex task of finding alternatives~\citep{he2021multi} or reverting to older versions~\citep{kula2018developers}. Societal reactions to protestware vary, with responses ranging from strong support to harsh criticism, and everything in between~\citep{massacci2022free}.
Through the GitHub comments and social media, we can also gain insights into the practicality of hosting protestware, and in what instances should protestware become tolerated. This, in turn, contributes to the creation of a more friendly software ecosystem.

In this paper, we study the disruptions caused by protestware and the immediate reactions it prompts.
Specifically, we explore the recent cases \caseOne~and \caseTwo~as instances of protestware.
\revise{We utilize \caseThree~and \caseFour~as baselines for the experiment. The first baseline, \caseThree, refers to a security vulnerability involving the hijacking of a maintainers account to release malicious versions of an npm package. This incident highlighted a significant attack involving high-profile cases involving the hijacking of legitimate developer accounts, including \caseThree.\footnote{\url{https://t.ly/pdVPa}} We selected \caseThree~because it arose during a period when protestware incidents were becoming prominent. 
The second baseline, \caseFour, also known as CVE-2021-44228 or Log4Shell, represents a critical security vulnerability. It significantly impacted the software ecosystem, affecting over 17,000 packages (around 4\% of the ecosystem) by December 19, 2021. Approximately 25\% of these affected packages had fixed versions available.\footnote{\url{https://security.googleblog.com/2021/12/understanding-impact-of-apache-log4j.html}} Log4Shell was assigned a maximum severity score of 10 on the Common Vulnerability Scoring System (CVSS) scale.\footnote{\url{https://logging.apache.org/log4j/2.x/security.html}} We chose \caseFour~due to its extensive impact and the rapid response from the open source community.}
To guide our study, we formulate the following two research questions:

%%%%%%%%%%%%%%%%%%%%%%%%%%%%%%%%%
\noindent
-- \RqOne\\
This question aims to quantify the spread of protestware discourse across the ecosystem and social media, determining both the speed and extent of discussions, which reflect community interest and concern. \\
\revise{\ul{Results} show that developers comparatively engage in protestware discussions similar to security vulnerabilities. We find that it takes seven days to reach 136 developers and generate 245 posts for \caseOne, and fourteen days to reach 155 developers and generate 720 posts for \caseTwo. Overall, protestware discussions spread faster compared to vulnerability discussions during the early stages of the interruption on the GitHub platform, while being slower on social media platforms like Hackernews and Reddit.}

\noindent
-- \RqTwo\\
     Qualitatively, we take a deeper investigation into the different narratives that revolve around the protestware, specifically focusing on the stances and communicative styles, themes emerging and mitigation strategies.
    \begin{itemize}
        \item \RqTwoA\\
            This question aims to measure how frequently developers express their stances on protestware and to identify the communicative styles they use in the discussions.\\
        \ul{Results} include a \revise{classification} of stances and communicative styles taken by developers. Among those who expressed a stance, ``opposing'' was more frequently associated with ``\revise{direct}'' and ``offensive/hate and toxic'' communicative styles, while ``favorable'' stances were often expressed with ``\revise{direct}'' styles.
        \item \RqTwoB\\
            With this research question, we aim to categorize community reactions to protestware. The analysis helps us understand the content of developer posts, highlighting thematic elements that provide insights into the diverse perspectives within the community.\\
            \ul{Results} include a taxonomy of themes that revolve around discussions of \textit{rights and ethics}, \textit{disseminate and response}, and \textit{reputation}.
             \item \RqTwoC\\
            This question aims to identify and analyze the suggestions provided by developers for mitigating the impact of protestware in software projects.\\
        \ul{Results} show a \revise{classification} of protestware mitigation. We found that 35\% of the posts provided technical instructions for mitigating protestware, with key strategies including dependency management (11\%), addressing license issues (7\%), and suggesting alternatives (3\%).
\end{itemize}

\noindent
-- \RqThree\\
      We ask this question to investigate the direct influence of protestware on developers' decisions. We aim to assess whether protestware leads to the abandonment or continued usage of dependencies, which reflects the perceived severity or acceptance of the protest.\\
    \ul{Results} show developers are less likely to abandon the protestware (i.e., out of 146 repositories that depend on \caseOne, only 15\% abandoned).

Insights from the study indicate how protestware discussions are multifaceted and diverse. 
Potential tool support includes early detection, distilling technical mitigation instructions, and potential to moderate toxic-side effects of protestware discussions. 
The study's contribution is the first large-scale exploration into protestware, presenting a taxonomy of narratives that drive this emerging phenomenon.

%%%%%%%%%%%%%%%%%%%%%%%%%%%%%%%%%

\section{Protestware and Vulnerabilities}
\label{sec:protestware_vulnerability}

\subsection{Case Studies}
Specifically, we focus on {two} prominent instances of protestware (color.js and es5-ext). 
We selected these two cases due to their significant engagement within the community. \caseOne, with over 3.3 billion downloads and dependency in more than 19,000 projects, has garnered substantial attention throughout its lifetime.\footnote{https://blog.sonatype.com/npm-libraries-colors-and-faker-sabotaged-in-protest-by-their-maintainer-what-to-do-now} Similarly, es5-ex maintains a consistent weekly download count exceeding million, underscoring its ongoing relevance and widespread use.\footnote{https://checkmarx.com/blog/new-protestware-found-lurking-in-highly-popular-npm-package/}
Although the \textit{node-ipc} and \textit{Faker.js} libraries are among the more infamous cases of protestware, these projects have been closed or replaced, with most historical information removed from both GitHub and the broader Internet. 
To fairly compare the disruption, we in addition introduce {two well-known security vulnerabilities (ua-parser and log4j) as the baseline}.

\begin{figure}[t]
    \centering
    \includegraphics[width=0.8\textwidth]{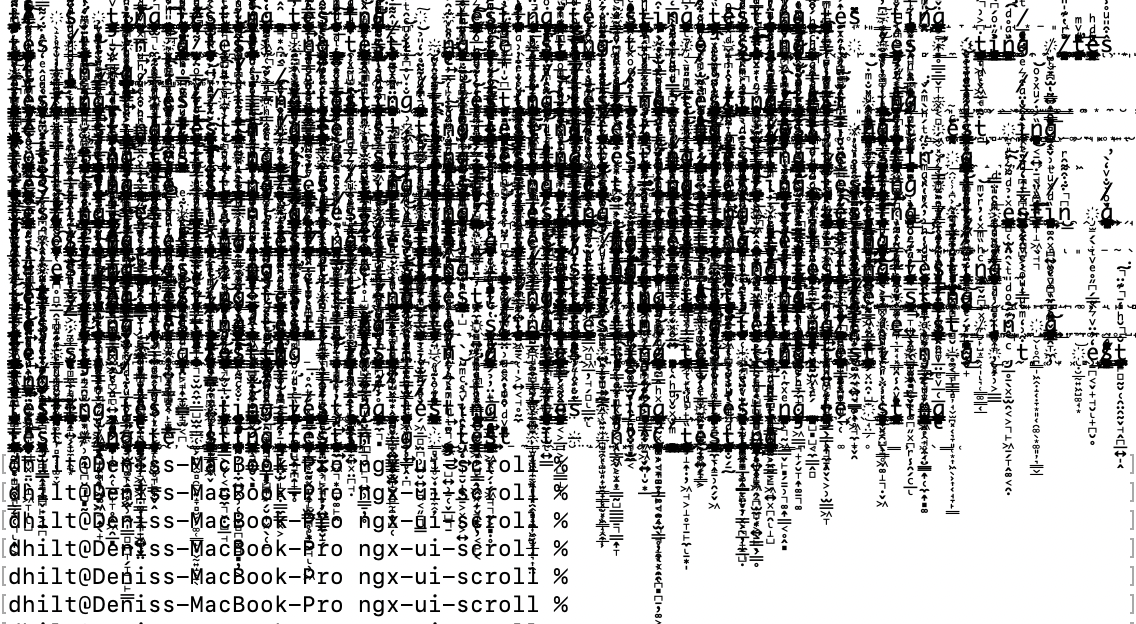}
    \caption{Output of protestware \caseOne}
    \label{fig:colors}
\end{figure}

\paragraph{\textbf{\caseOne~- Protest Against Fortune 500 Companies}}
Marak's motivations were explained on his personal blog from April 2021, entitled ``Monetizing Open-source is problematic''. He became well-known for his statement on the now-banned repository, Faker.js:
\begin{formal}
    \faGithub~\color{coolblack}{\textit{``No more free work from Marak \\- Pay Me or Fork This''}}
\end{formal}

The reaction from the open-source community has been diverse, sparking news articles and discussions on protestware. 
Marak's actions have stirred debates on many fronts, from the need for testing to the rights and responsibilities of code owners.
As shown in Figure \ref{fig:colors}, Marak also turned his \caseOne~library into protestware, leading to GitHub banning his action. A subsequent series of events opened a discussion on whether Marak's decision to self-sabotage his library was justified.
\begin{formal}
\color{coolblack}{
\faReddit~\textit{{What? I'm totally on his side! Open-source maintainers should be supported by the companies who use their work. I hope they work out proper licensing deals, as suggested in the video.}}}
\end{formal}

\paragraph{\textbf{\caseTwo~- Protest against the Ukraine War}}
Similarly, the popular NPM package \caseTwo, with nearly 13 million weekly downloads, was updated to include a ``Call for peace'' message targeting machines configured to Russian time zones. 
Snippets of the code\footnote{\url{https://github.com/medikoo/\caseTwo/blob/main/_postinstall.js}} are shown in Listing 1, which is only claimed to be invoked during testing.
The maintainer's thoughts are captured in the following quote:
\begin{formal}
\faGithub
\color{coolblack}{
    \textit{I'm opening this issue to serve as a single place to serve any (possibly constructive) discussions, towards the anti-Russian invasion on Ukraine manifest that's part of this package. As I mentioned in 28de285 whole point of this manifest, is to urge all Russians that see it, to seek reliable sources of information of what's going on, as you won't get it from official media sources in Russia.
    It's very important for you and the whole world that you realize what's really going on. Note that I'm affected by the situation directly, and it's why I've decided to put this manifest in...}}
\end{formal}

\begin{lstlisting}[language=Python,
caption={Protestware snippet to identify user location},
label=code:userhome]
// Broadcasts "Call for peace" message when package is installed in Russia, otherwise no-op
...
try {
	if (
		[
			"Asia/Anadyr", "Asia/Barnaul", "Asia/Chita", "Asia/Irkutsk", "Asia/Kamchatka",
			"Asia/Khandyga", "Asia/Krasnoyarsk", "Asia/Magadan", "Asia/Novokuznetsk",
			"Asia/Novosibirsk", "Asia/Omsk", "Asia/Sakhalin", "Asia/Srednekolymsk", "Asia/Tomsk",
			"Asia/Ust-Nera", "Asia/Vladivostok", "Asia/Yakutsk", "Asia/Yekaterinburg",
			"Europe/Astrakhan", "Europe/Kaliningrad", "Europe/Kirov", "Europe/Moscow",
			"Europe/Samara", "Europe/Saratov", "Europe/Simferopol", "Europe/Ulyanovsk",
			"Europe/Volgograd", "W-SU"
		].indexOf(new Intl.DateTimeFormat().resolvedOptions().timeZone) === -1
....
\end{lstlisting}

\section{Data Collection}
\label{sec:data_collection}
In this section, we will describe the process of data preparation from both GitHub and the social media platform.

\begin{figure}[]
  \centering
  \includegraphics[width=1\linewidth]{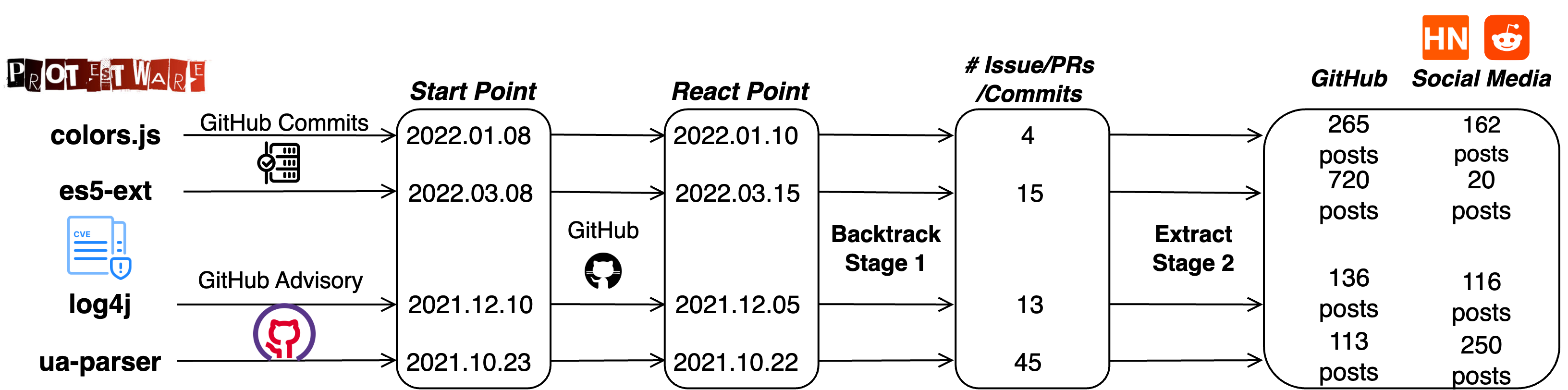}
  \caption{An Overview of the Data Collection Process}
  \label{fig:data_collect}
\end{figure}

\paragraph{\textbf{Protestware start and react points.}}
To capture all protestware discussions and their reactions, we first needed to identify the \textit{start point} of each event. Hence, we looked at the actual commit where the protester inserted the protest code into the code base. As shown in Fig.~\ref{fig:data_collect}, the code containing the protestware was committed on 2022.01.08 for colors.js and 2022.03.08 for es5-ext. For our baselines, we used the GitHub advisory's reported date as the start point for that vulnerability (2021.10.23 for \caseThree~and 2021.12.10 for \caseFour).\\
In terms of the \textit{reaction points}, we searched for the first issue that was raised in the protestware repository.
As such, we identified the following issues as the reactions from GitHub for all four studied cases:
(i.e., 2022.01.10 for \caseOne\footnote{\url{https://github.com/Marak/\caseOne/issues/289}}, 
2022.03.15 for \caseTwo\footnote{\url{https://github.com/medikoo/\caseTwo/issues/116}}, 
2021.12.05 for \caseFour\footnote{\url{https://github.com/apache/logging-log4j2/pull/608}}, 
2021.10.22 for \caseThree\footnote{\url{https://github.com/faisalman/ua-parser-js/issues/536}}).

\ul{\textit{Recursive Backtracking}}.
To measure the spread of the protestware reactions throughout the rest of GitHub and social media, we implemented a recursive backtracking algorithm to extract any links (i.e., Issues, PRs, and commits). Note that in this paper, we refer to all conversations inside issues, PRs and commits as posts.
We used the GitHub Rest API\footnote{\url{https://docs.github.com/en/rest?apiVersion=2022-11-28}} to retrieve corresponding metadata including the post body.
In compliance with GitHub's download rate limitations, we systematically downloaded the data over a period from 24th April to 30th June, 2023.
Our backtracking followed two stages:

\textbf{Backtrack Stage 1} - Recursive collection of all GitHub links. We applied a regular expression to capture all links, e.g., 
\texttt{https://github.com/([$\wedge$/]+)/([$\wedge$/]+)}
\texttt{/issues/([$\textbackslash$d]+)} for identifying any hyperlinks (similarily for PRs and commits) in the posts from the protester's repositories.
As shown in Fig~\ref{fig:data_collect}, we obtained four links for the Marak case (two issues and two commits across five repositories), three links for \caseTwo~ case (two issues and one commit across two repositories), 13 links for \caseFour~case (5 PRs and 8 commits across one repository), and 26 links for \caseThree~case (14 issues, 8 PRs, and one commit across four repositories).

\revise{\textbf{Extract Stage 2} - In this stage, we cross-reference mentions in all backtracked posts by collecting a series of events from GitHub links collected in Backtrack Stage 1 . GitHub provides this information, identifying any issues, pull requests (PRs), and commits that were referenced or cross-referenced~\citep{reference, crossreference} from another issue, pull request, or commit within the timeline. All collected posts are depicted in Fig~\ref{fig:data_collect}.
}
\paragraph{\textbf{Discussions in Social Media.}}
To measure the spread in other outlets such as social media, we chose to focus on Reddit and Hacker News, since these {two} media platforms are popular for developers to submit posts related to software development and they have been commonly studied in the context of software engineering research~\citep{aniche2018modern}.
To remove noise from non-developers and ensure data quality, we input the issues, PRs, and commits referenced as the start point for the protestware.
Note that in our study, the comment and its reply are regarded as two distinct posts.
We retrieved three threads from Reddit and one thread from Hacker News.
To collect their posts, we utilized various APIs, such as the Python Reddit API Wrapper for Reddit,\footnote{\url{https://github.com/praw-dev/praw}} and the Request library for Hacker News. 
As shown in Fig.~\ref{fig:data_collect}, we ended up with 540 posts.

\section{Empirical Results}
\label{sec:empirical_results}
We now present the approaches and the results of our proposed research questions.
\subsection{Impact of Disruption (RQ1)}
To address RQ1, we answer the question as shown below.

\paragraph{\ul{\textbf{Spread throughout the Ecosystem (RQ1)}}}

\revise{Inspired by related work~\citep{kanuri2018scheduling}, we defined social media metrics to measure the engagement and how quickly discussions of protestware spread within the GitHub ecosystem and the external social media.
We use the four origin issues as the starting point of each studied case and use the following definitions:
\begin{itemize}
    \item \textit{Post.} All conversations are regarded as posts for any issue, PR, commit, and social media posts. Each post has a timestamp, the post context in the text, and the developer who is the author of the post.
    \item \textit{Engaged developer:} This is a count of unique developers who have authored any protestware posts after the start point. Hence, we only count the first instance of the engaged developer post.
    \item \textit{Engaged post:} This is the count of unique posts from the start point of the protestware. Similar to ``engaged developers'', this includes only posts created after the start point, removing duplicates.
\end{itemize}
Based on the above definitions, we now introduce our engagement metrics.
\begin{itemize}
    \item \textit{Developer engagement metric. }This measures the number of developers engaged within one, seven, and fourteen days from the start point of the disruption.
    
    \item \textit{Post engagement metric. }This measures the number of posts generated within one, seven, and fourteen days from the start point of the disruption. 
    
    \item \textit{Social media post engagement metric (Hackernews and Reddit). }This measures the number of posts made on social media platforms (Hackernews and Reddit) within one, seven, and fourteen days from the start point of the disruption.
\end{itemize}
}

%%%%%%%%%%%%%%%%%%%%%%%%%%%%%%%%%%
\begin{table*}[]
\centering
 \caption{Summary Statistics for Time-to-Engage (RQ1)}
 \label{tab:engagement}
 \begin{threeparttable}
\begin{tabular}{lrrrr}
\toprule
 & \caseOne & \caseTwo & {\caseFour} & \caseThree \\ \midrule
{\textbf{Developer Engagement Metric} \faGithub}                             &       \\
\hspace{1em}- one day  & 9 devs & - & \fcolorbox{navy}{light}{24 devs} & 21 devs\\
\hspace{1em}- seven days & \fcolorbox{navy}{light}{136 devs} & 37 devs & 55 devs&  50 devs \\ 
\hspace{1em}- fourteen days & 139 devs & \fcolorbox{navy}{light}{155 devs} & 56 devs &  56 devs \\ 
 \midrule
{\textbf{Post Engagement Metric \faGithub}}                             &       \\
\hspace{1em}- one day & 11 posts & -& \fcolorbox{navy}{light}{32 posts} &   24 posts\\
\hspace{1em}- seven days & \fcolorbox{navy}{light}{245 posts} & 148 posts & 76 posts &  70 posts \\
\hspace{1em}- fourteen days & 250 posts & \fcolorbox{navy}{light}{720 posts} & 77 posts & 77 posts   \\ 

 \midrule
{\textbf{Hackernews Post Engagement Metric \faHackerNews}}                      &     \\
\hspace{1em}- one day & 7 posts & -& \fcolorbox{navy}{light}{19 posts} &  7 posts \\
\hspace{1em}- seven days & 12 posts & - & \fcolorbox{navy}{light}{41 posts} &  7 posts \\
\hspace{1em}- fourteen days & 12 posts & - & \fcolorbox{navy}{light}{41 posts} & 7 posts   \\ 

\midrule
{\textbf{Reddit Post Engagement Metric \faReddit}}                      &     \\
\hspace{1em}- one day & 53 posts & -& 2 posts & \fcolorbox{navy}{light}{ 146 posts} \\
\hspace{1em}- seven days & 129 posts & - & 50 posts & \fcolorbox{navy}{light}{191 posts}  \\ 
\hspace{1em}- fourteen days & 130 posts & 1 post & 51 posts &  \fcolorbox{navy}{light}{191 posts} \\

\bottomrule
\end{tabular}
\begin{tablenotes}
   \item \textit{devs = developers}
   \item \textit{\textbf{Note}: The highlights represent the largest number of developers and posts that the protestware and vulnerabilities reached the predetermined engagement metrics.}
\end{tablenotes}
\end{threeparttable}
\end{table*}
%%%%%%%%%%%%%%%%%%%%%%%%%%%%%%%%%%

\revise{\ul{\textbf{Discussions of protestware spread like wildfire on GitHub, but not on social media.}}
Table~\ref{tab:engagement} summarizes the engagement metrics. Our analysis shows that while protestware discussions spread quickly on the GitHub platform, security vulnerabilities tend to spread faster on social media. For example, on GitHub, \caseOne~engaged 136 developers within seven days, whereas \caseThree~engaged only 50 developers at the same time. However, \caseThree~generated 146 posts in just one day, illustrating the faster dissemination of vulnerability discussions compared to protestware on social media platforms.}

\begin{tcolorbox}[colback=gray!5,colframe=awesome,title= RQ1 Summary]
\revise{
Our quantitative results indicate that discussions about protestware (i.e., \caseOne~and \caseTwo) tend to spread more quickly on the GitHub platform, while security vulnerabilities (i.e., \caseThree~and \caseFour) engage faster on social media.\ For example, within the first seven days, \caseOne~engaged 136 developers and generated 245 posts, demonstrating its significant impact on the GitHub ecosystem. In contrast, security vulnerabilities like \caseThree~spread faster on social media, with 146 posts engaged in just one day. }
\end{tcolorbox}

\subsection{Characteristics of Posts (RQ2)}
%%%%%%%%%%%%%%%%%%%%%%%%%%%%%%%%%%%%%%%%%%%%%%%%%%%%%%%%%%%%%%%%%%%%%%%%%%%%
\begin{table}[]
\centering
\caption{Dataset to answer the Qualitative Analysis in RQ2} 
\label{tab:RQ2}
\begin{tabular}{lrr}
\toprule
 & \caseOne & \caseTwo \\ \midrule
\textbf{From GitHub}&&      \\
% \hspace{1em}- \# Issues/PRs/Commits from Recursive Backtracking Stage 1 & 3 & 2 \\
\hspace{1em}- \# protestware repo posts (1,458) & 797 & 661 \\ 
\midrule
 \textbf{From Social Media }&                              \\
\hspace{1em}- \# posts from RQ1 ({182}) & 162 & 20 \\
\midrule

\textbf{After Filtering }& \\ 
\hspace{1em}- Sum \# posts for RQ2 (GitHub + Social Media) &  578 & 106 \\ \bottomrule
\end{tabular}
\end{table}
%%%%%%%%%%%%%%%%%%%%%%%%%%%%%%%%%%%%%%%%%%%%%%%%%%%%%%%%%%%%%%%%%%%%%%%%%%%
To address RQ2, we answer three sub-research questions by classifying the topics of discussion in the conversations about protestware. We conducted an in-depth assessment of developers' positions and the nature of their posts. \revise{Due to the complex and nuanced nature of protestware descriptions in natural language \citep{cheong2023ethical, kula2022war}, it is challenging to use any existing tools for detection.
Hence, we opted for a manual coding approach. }
\revise{Note that for RQ2, since we focus only on protestware posts, we exclude the baseline cases (i.e., \caseThree~and \caseFour) from our analysis. Additionally, to reduce noise in the dataset, we filtered the data by including only posts that contain links directly related to \caseOne~and \caseTwo.}
% Note that for RQ2, since we focus only on protestware posts, we exclude the baseline from our analysis (i.e., \caseThree~and \caseFour) 

\paragraph{\ul{\textbf{Qualitative Dataset.}}}

Table \ref{tab:RQ2} shows that our dataset contains a total of 1,640 posts (1,458 from GitHub and 182 from social media). Following prior works~\citep{hata20199, wang2021understanding}, we adopted a systematic manual coding approach with multiple rounds to identify the relevant posts. 
Since the quantitative analysis in RQ1 is an overestimation, especially with backtracking the mentions in our approach, we focus only on the results of the recursive backtracking from stage 1.

\paragraph{\ul{\textbf{Emerging themes.}}}
Since manual analysis of the posts is a tedious approach, we adopted a systematic but iterative process. 
For all sub-research questions, we then took a holistic approach to conduct the classifications with the three different dimensions in mind (i.e., stances and communicative styles, themes, and mitigation strategies).
Our iterative process is broken into three processes. 
Note that since combining all sub-questions, each \revise{classification} was derived independently. 

\begin{itemize}
    \item \textit{Initial Coding}. The goal of the initial coding was to ensure we had saturation and coverage of all themes. Initially, two researchers reviewed 20 posts together and discussed potential categories and codes, leading to the development of a preliminary code schema. Based on this schema, three researchers collaboratively coded 30 posts, achieving Kappa values of 0.73 for technical information, 0.80 for dissemination and response, 0.75 for stance, 0.67 for reputation, 0.67 for communicative style, and 0.78 for rights and ethics. These agreements were considered moderate \citep{viera2005understanding}. The researchers then distributed the rest of the dataset among themselves.
    \item \textit{Round-table discussions for consensus}. After completing the individual coding, the researchers reconvened to discuss and resolve all highlighted uncertainties, resulting in a total of 139 posts being reviewed. The three researchers sat together to discuss and reach a consensus on all codes, especially in cases where the coders were not confident. After extensive discussion, the three researchers decided on each classification.
    \item \textit{Final consensus and definitions refinement}. After all classifications were made, the three researchers then carefully refined the definitions of each code to ensure that the themes were properly represented. 
\end{itemize}

\begin{figure}[]
    \centering
    \includegraphics[width=0.8\linewidth]{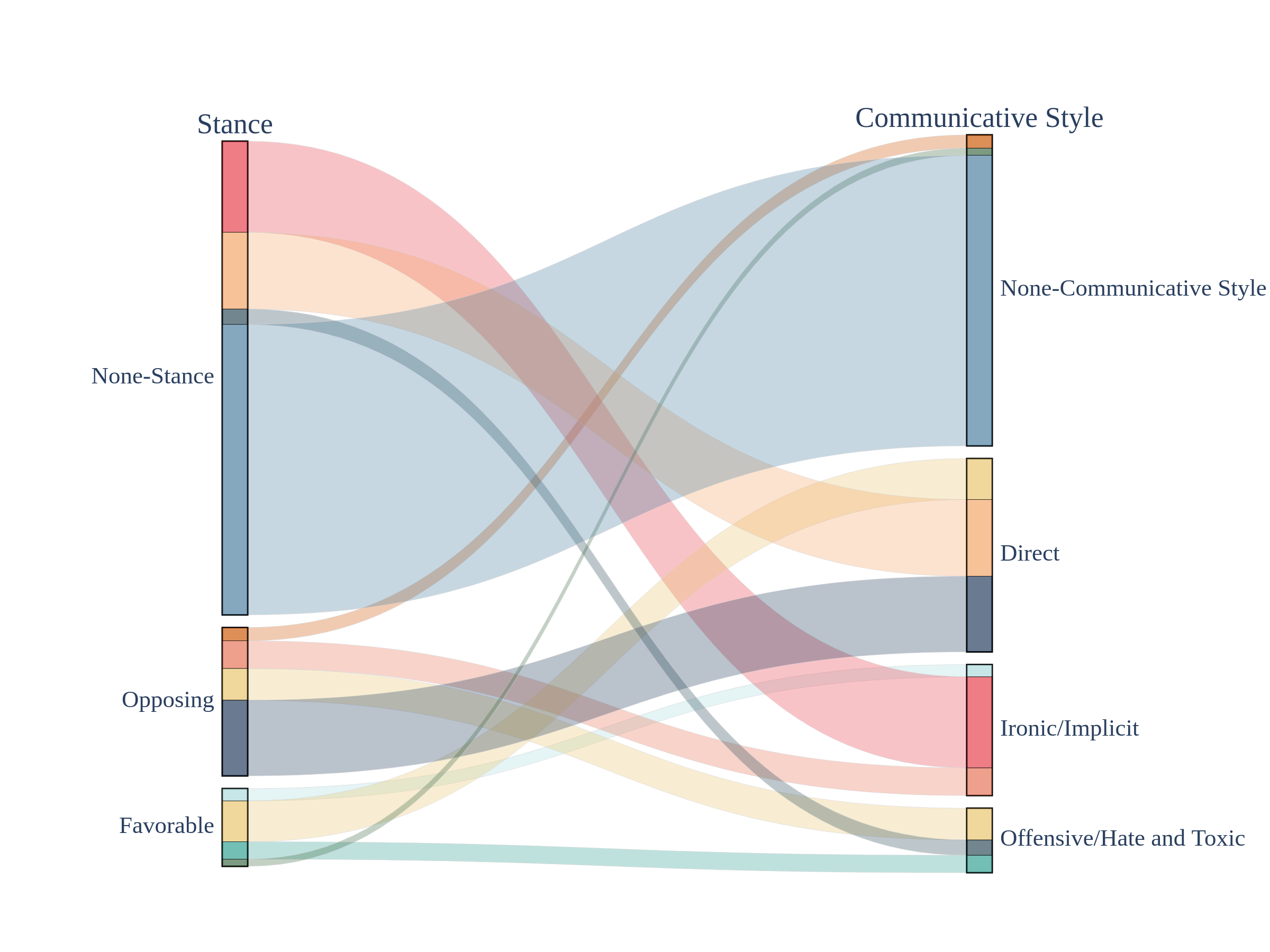}
    \caption{Distribution between the stance and communicative style.}
    \label{fig:relationship_stance_sentiment}
\end{figure}

\ul{\textbf{Developers were primarily opposing the protestware, using either direct or offensive/hate and toxic communicative styles (RQ2a)}}
\revise{The first dimension of protestware discussions is classified by both stance and communicative style.}
As shown in Fig. \ref{fig:relationship_stance_sentiment}, where we applied co-occurrence analysis, most developers in the discussions did not express a clear stance (68\%). Among the 32\% who did, the majority were opposed to protestware (21\%), with their communication often being \revise{direct}, offensive, or characterized by hate and toxicity (9\%). This indicates a polarized and frequently harsh discussion environment surrounding protestware issues. \revise{To assess the significance of the relationship between communicative styles and stances, we applied Pearson's Chi-Squared test \citep{pearson1900x}, which is commonly used to evaluate relationships between categorical variables~\citep{wang2023more}. The test confirmed a significant relationship between communicative styles and stances, with a p-value of $<$ 0.001, indicating that the association between communicative styles and stances is highly significant.}
We now provide the definitions and examples for each of our codes as follows:
\begin{itemize}
    \item \textbf{Stance}
        \begin{itemize}
            \item \textit{Favorable (11\%).} The post expresses support for the protester and the actions of the protestware, for example ``Creative choices like this (while disruptive) do keep us alert. I'm sympathetic to all of us who this impacted, but I'm in support of playfulness, developer wellbeing and personal agency for contributors and maintainers such as @Marak''.
            \item \textit{Opposing (21\%).} The post expresses opposition to the protester and the actions of the protestware, for example ``There are 44 contributors to this project, it is their code, not his.''
            \item \textit{None (68\%).} The post did not express support or opposition to the protestware, or it is hard to tell, for example ``Ah isn't that related to fakerjs? Wasn't it maintained by the same person or?''
            
        \end{itemize}
\end{itemize}

\begin{figure}[t]
    \centering
    \includegraphics[width=0.8\linewidth]{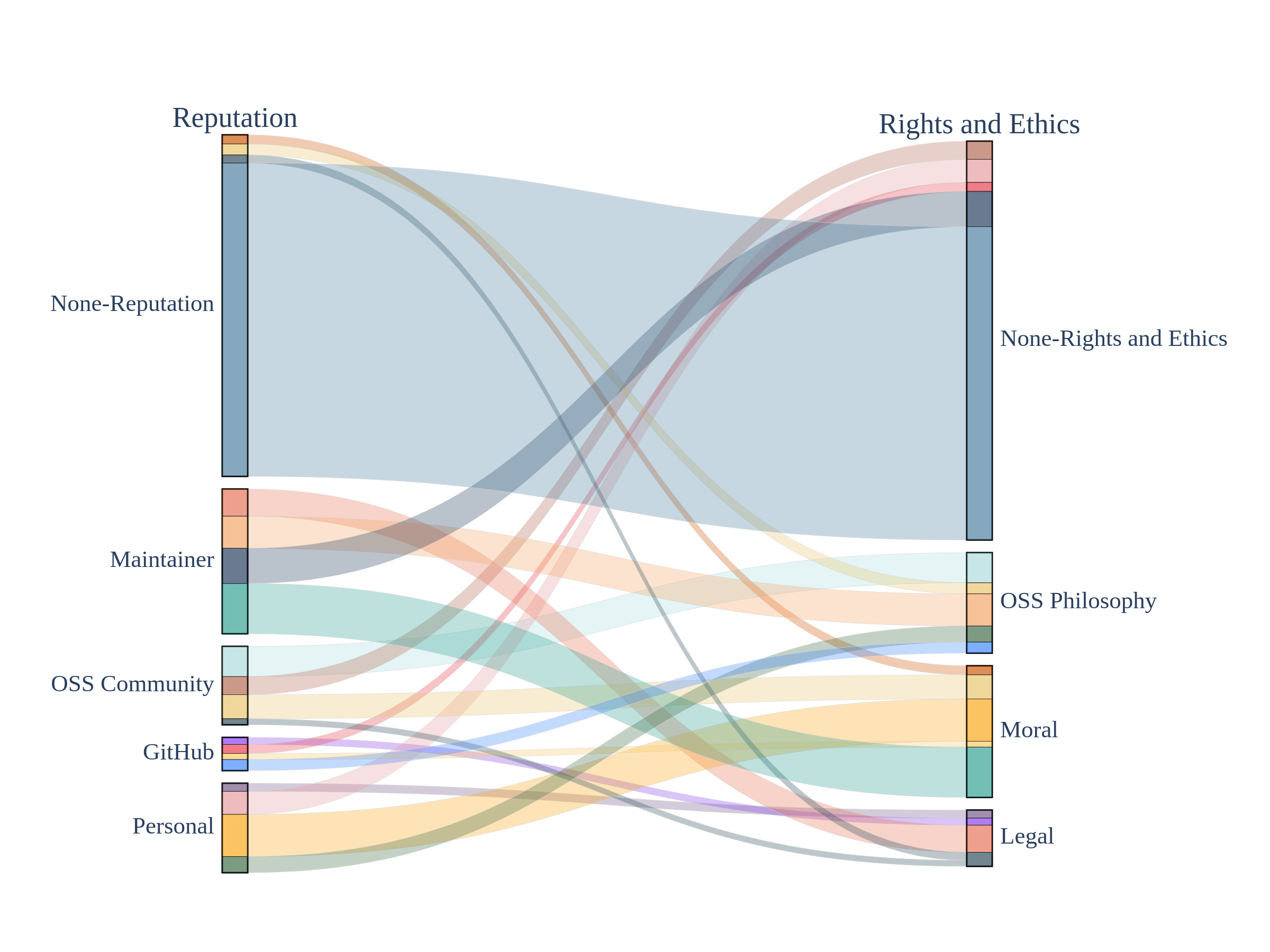}
    \caption{Distribution of the different reputations and their rights and ethics expressed. }
    \label{fig:repu_right}
\end{figure}

\begin{itemize}
    \item \textbf{Communicative Style}
        \begin{itemize}
            \item \textit{Offensive/Hate and Toxic (9\%).} The post contains insults or profanity, or the post is characterized by malicious intent, for example ``Wow, another NPM shitshow. The major problem is why the other libraries don't fix their dependency versions.''
            \item \textit{Ironic/Implicit (19\%).} The post contains expressions that have opposite literal meanings to real meanings. The implicit nature refers to the post containing anecdotal narratives, for example ``v1.4.44-liberty-2 is working great for me, @Marak. Thanks for pushing this update quickly, my console has never looked better.''
            \item \textit{\revise{Direct} (28\%).} \revise{The post contains expressions that take a firm and conclusive stance on an issue, clearly stating an opinion with no alternative meanings or interpretations, }for example ``He published malicious code when he has $>20$ million weekly downloads, not acceptable.''
            \item \textit{None (44\%).} The post does not express any sentiments in terms of offensive, hate and toxic, ironic, implicit, explicit or \revise{direct}, for example ``Folks interested in what's going on with colors.js should have a look at the update from DABH (who had made the more recent and substantial changes) on GitHub. \url{https://t.co/uMlIAjqiko}''
            
        \end{itemize}
\end{itemize}
%%%%%%%%%%%%%%%%%%%%%%%%%%%%%%%%%%%%%%%%%%%%%%%%%%%%%%%%%%%%%%%%%%%%%%%%%%%%
\paragraph{\ul{\textbf{Topics of Discussion (RQ2b)}}} 
\revise{The second dimension, as shown in Fig \ref{fig:overviewOpinions} is a taxonomy of the topics discussed in relation to protestware.} As shown in Fig \ref{fig:repu_right}, the topic highlights the different focuses and concerns developers have when discussing protestware, particularly in terms of reputation and rights and ethics, the detailed categories are as follows:

\begin{figure}[t]
    \centering
    \includegraphics[width=1\linewidth]{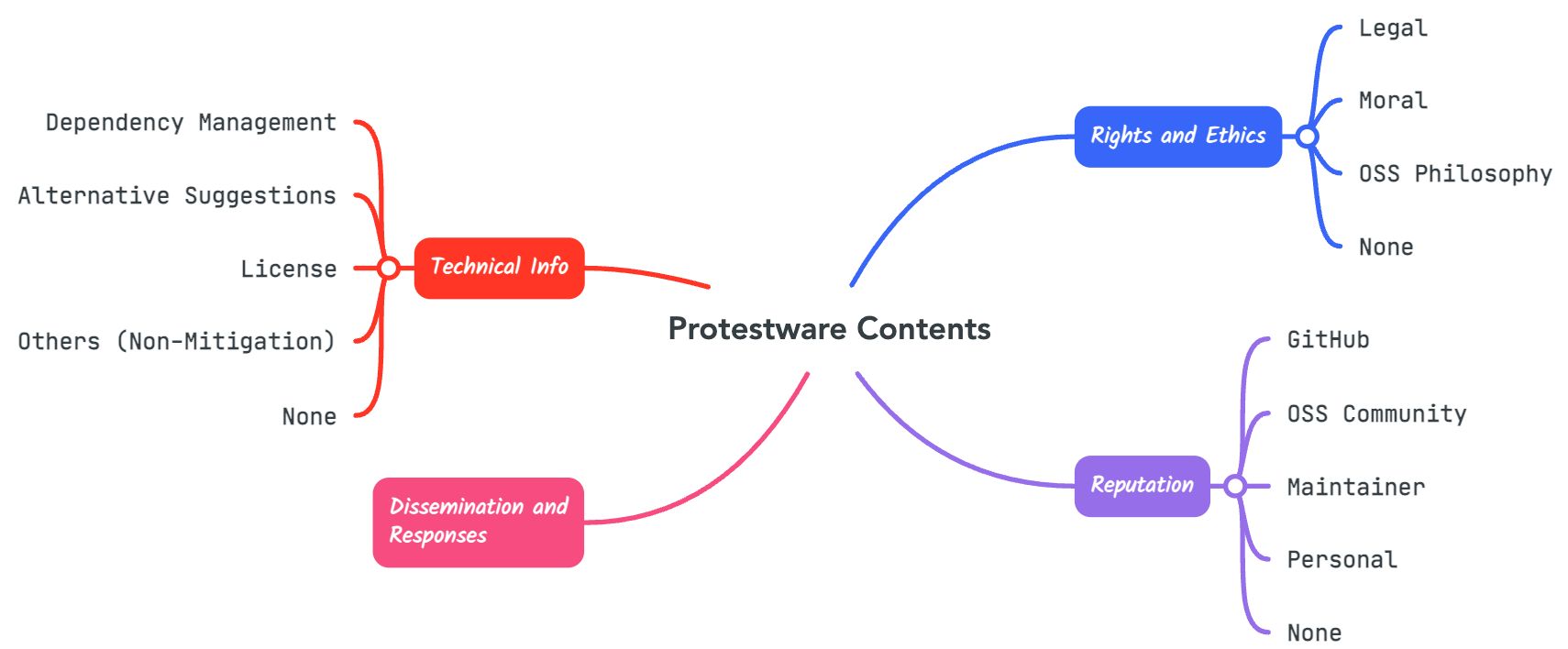}
    \caption{A mind map of the themes emerging from protestware discussions}
    \label{fig:overviewOpinions}
\end{figure}

\begin{itemize}
    \item \textbf{Rights and Ethics}
        \begin{itemize}
            \item \textit{Legal (8\%).} The post owner expresses the legal implications of the protestware, for example ``He's using a MIT license, which grants the right for people to use his work commercially, for free.''
            \item \textit{Moral (19\%).} The post owner expresses ethical considerations and consequences of the protestware, for example ``There are 44 contributors to this project, it is their code, not his.''
            \item \textit{OSS Philosophy (15\%).} The post expresses views that are related to the ethical philosophy behind open source software, for example ``Also known as ``contribute back to the community or pay up'' :laughing: ''
            \item \textit{None (58\%).} The post does not express any rights or ethical considerations regarding the protestware, for example ``github and npm are two different things - you can publish to npm without ever pushing the code to github''
        \end{itemize}
\end{itemize}

\begin{itemize}
    \item \textbf{Disseminate and Response}
        \begin{itemize}
            \item \textit{True (21\%).} The post contains external links of information related to the protestware. It does not contain any views of the post owner, for example ``it's no bug, they vandalized their own project: \url{https://github.com/Marak/colors.js/commit/074a0f8ed0c31c35d13d28632bd8a049ff136fb6 }''
            \item \textit{False (79\%).} The post does not contain any external links that are information on the protestware, for example ``Ah isn't that related to fakerjs? Wasn't it maintained by the same person or?''
        \end{itemize}
\end{itemize}
%%%%%%%%%%%%%%%%%%%%%%%%%%%%%%%%%%%%%%%%%%%%%%%%%%%%%%%%%%%%%%%%%%%%%%%%%%%%

%%%%%%%%%%%%%%%%%%%%%%%%%%%%%%%%%%%%%%%%%%%%%%%%%%%%%%%%%%
%%%%%%%%%%%%%%%%%%%%%%%%%%%%%%%%%%%%%%%%%%%%%%%%%%%%%%%%%%
\begin{table}[t]
\centering
\caption{Distribution of posts with Technical Information from RQ2c}
\label{tab:technical_related_posts}
\begin{tabular}{lrr}
\toprule
& \caseOne & \caseTwo  \\ \midrule
With Technical Info &  &   \\ 
 \hspace{2em} - Other & 84 (12\%) & 8 (1\%)  \\
 \hspace{2em} - Dependency Management & 77 (11\%) & -  \\ 
\hspace{2em} - License & 46 (7\%) & 1 (0\%)  \\
\hspace{2em} - Alternative Suggestions & 23 (3\%) & -  \\ 
\midrule
 Sum &\textbf{230} (34\%)& \textbf{9} (1\%)\\
\bottomrule
\end{tabular}
\end{table}

%%%%%%%%%%%%%%%%%%%%%%%%%%%%%%%%%%%%%%%%%%%%%%%%%%%%%%%%%%

\begin{itemize}
    \item \textbf{Reputation}
        \begin{itemize}
            \item \textit{Maintainer (21\%).} The post is directed to the protester's reputation as a maintainer in the OSS community, for example ``That can happen too: Maintainer of a popular \#javascript library colors.js intentionally pushed an infinite loop into the code, crashing thousands of other libraries that depends on it. here's the malicious commit: \url {https://github.com/Marak/colors.js/commit/074a0f8ed0c31c35d13d28632bd8a049ff136fb6} \#javascriptdeveloper''
            \item \textit{OSS Community (11\%).} The post is directed to the OSS community reputation that is affected by the protestware, for example ``Wow, another NPM shitshow. The major problem is why the other libraries don't fix their dependency versions.''
            \item \textit{Personal (13\%).} The post is a personal attack on the protester reputation. Not on their role as a maintainer, for example ``Also.. just a note.. colors.js actually had co-maintainers. He removed their access so they couldn't fix things. That's an extra level of cringe for me.''
            \item \textit{GitHub (5\%).} The post is directed to the reputation of the  GitHub Platform, for example ``Github's responsibility should have been to treat him like they treat everyone else and when they removed him from the platform they removed his content and code.''
              \item \textit{None (50\%).} The post is not related to reputation, for example ``Color contains maleware as far as I know? Doesn't it?''
        \end{itemize}
\ul{\textbf{Developers provide technical information that sometimes provides mitigation instructions (RQ2c)}}
\revise{The third dimension involves the classification of mitigation strategies.}
As shown in Table \ref{tab:technical_related_posts}, we classified technical posts into two categories: those that provide mitigation instructions and those that do not. 
Developers are more likely to provide mitigation strategies for \caseOne~compared to \caseTwo.
After discussion, we then were able to code the technical posts as the following:
\begin{itemize}
    \item \textit{With Technical Instructions}
        \begin{itemize}
            \item \textit{Alternative Suggestions (3\%).} The post contains an alternative fix to the protestware, for example ``For developers looking to migrate, alternatives of colors are: chalk, kleur, cli-color, colorette.''
            \item \textit{License (7\%).} The post contains some discussion on the licensing to mitigate the protestware, for example `` He's using a MIT license, which grants the right for people to use his work commercially, for free.''
            \item \textit{Dependency Management (11\%).} The post suggests changing (upgrading/downgrading) dependencies to mitigate the protestware, for example ``v1.4.1 is also broken v1.4.0 is working.''
            \item \textit{Other (13\%).} The post mentioned other methods to mitigate protestware that are not mentioned above, for example ``I can't launch because of this issue. Assuming this never gets fixed, what's the proper way to force npm to use an older version of this dependency for `http-server'? Edit the lock file?''
        \end{itemize}
    \item \textit{Without Technical Instructions }
        \begin{itemize}
            \item \textit{None (65\%).} The post does not contain any technical suggestions, for example ``Ah isn't that related to fakerjs? Wasn't it maintained by the same person or?''
        \end{itemize}
\end{itemize}
        
\end{itemize}

\begin{tcolorbox}[colback=gray!5,colframe=awesome,title= RQ2 Summary]
Our analysis reveals the diverse stances, communicative styles, discussion topics, and mitigation strategies that developers use when discussing protestware.
\begin{itemize}
    \item Answering \textbf{RQ2a}, most developers (68\%) did not express a clear stance, but those who did were primarily opposed to protestware (21\%), using \revise{direct} (28\%) or offensive/hate and toxic (9\%) communicative styles.
    \item Answering \textbf{RQ2b}, developers' discussions on protestware centered around the topics of moral (19\%), OSS philosophy (15\%), legal (8\%), dissemination and response (21\%), and reputation about maintainer (21\%).
    \item Answering \textbf{RQ2c}, developers provided various mitigation strategies for protestware, including dependency management (11\%), license discussions (7\%), alternative suggestions (3\%), and other technical instructions (13\%), although 65\% of posts lacked technical suggestions.
\end{itemize}
\end{tcolorbox}

\subsection{Abandoning the Protestware (RQ3)}

To understand the influence of protestware, we conducted a quantitative analysis to investigate whether developers who discussed protestware ended up abandoning the associated dependency. We analyzed all the repositories from our collected dataset, focusing on those that (i) are valid JavaScript repositories, and (ii) use protestware as a dependency. To achieve this, we adopted an approach similar to prior works \citep{Cogo2019,decan2017empirical,wittern2016look}, identifying projects that contain the \texttt{package.json} file. Within these JSON files, we then determined whether or not protestware was documented as a dependency. Our analysis revealed that 4,961, 1, and 5,449 repositories met the above criteria for \caseOne, \caseTwo, and \caseThree. As shown in the Table \ref{tab:abandoned}~all discussions for \caseTwo~come from a single repository. This is due to the stage one backtracking only identifying one repository. We confirmed that this was the \caseTwo~repository itself. To detect whether or not a repository abandoned the protestware, we use the PyDriller package\footnote{\url{https://github.com/ishepard/pydriller}} to mine the historical dependency changes, from the date of the disruption until the latest commit in our dataset. 
Inspired by the work of \citet{TSECongruence22}, we focus on the following four types of dependency changes: dependency add, remove, upgrade, and downgrade.
In addition, we measure the popularity of the influenced repositories. 
Specifically, we employ the GitHub API to retrieve common popularity metrics, i.e., \#Forks, \#Commits, \#Stars, and \#Contributions~\citep{borges2016understanding}.

Since Log4j is not written in JavaScript, we employed a separate approach. We used the backtrack stage 1 method to identify all related GitHub projects. We then filtered for projects primarily using Java and retrieved their dependency lists via the GitHub GraphQL API. This allowed us to check if these projects still use Log4j and determine the version in use.\\

\begin{table*}[]
\centering
\caption{Influenced Repositories that Abandoned the Protestware (RQ3)}
\label{tab:abandoned}
  \resizebox{\textwidth}{!}{
\begin{tabular}{lcc|cc}
\toprule
 & \caseOne & \caseTwo & \caseThree  &\caseFour \\ \midrule
  \textbf{Collected Dataset} &           &       \\ 
\hspace{1em}- \# all collected repositories & 17,457 & 1 & 17,280  & 20,900\\ 
\textbf{Backtrack Stage 1 Repositories} &           &       \\
\hspace{1em}- \# same language Repositories         & 4,961 & 1 & 5,449 & 512  \\
\hspace{1em}- \# Not Dependent & 4,815 & 0 & 5,431  & 483\\\midrule
\hspace{1em}- \textbf{\# Dependent Repos}  & \textbf{146} & \textbf{1}  & \textbf{18} &\textbf{29} \\ \midrule
 \textbf{Repo Characteristics}          &     \\
\hspace{1em}- \# Forks (Median-Max) & Med=46, Max=14,773 &  915 & Med=142, Max=3,045 & Med=25, Max=15,176\\
\hspace{1em}- \# Commits (Median-Max) & Med=604, Max=35,552 & 9,670 & Med=1,974, Max=35,627 & Med=442, Max=11,495\\
\hspace{1em}- \# Stars (Median-Max) & \fcolorbox{navy}{light}{Med=124, Max=108,008} & 2,170 & \fcolorbox{navy}{light}{Med=445,Max=25,489} & \fcolorbox{navy}{light}{Med=46, Max=32,990}\\
\hspace{1em}- \# Contributors (Median-Max) &  Med=23, Max=429 & 358 & Med=22,Max=361 & Med=11, Max=30\\\midrule
\textbf{Detected Dependency Changes}          &     \\
\hspace{1em}- No change (\%)  & \fcolorbox{navy}{light}{77\%} & \fcolorbox{navy}{light}{100\%} & 6\%& 3\% \\
\hspace{1em}- \faPlus~added Protestware   (\%) & 0.5\% & - & 6\% & -\\
\hspace{1em}- \faRemove~removed Protestware  (\%) & 15\% & - & 6\% & 6\%\\
\hspace{1em}- \faAngleDoubleUp~upgraded Protestware  (\%) & 7\% & - &  \fcolorbox{navy}{light}{82\%} & \fcolorbox{navy}{light}{54\%} \\
\hspace{1em}- \faAngleDoubleDown~downgraded Protestware  (\%) & 0.5\% &  - & - & -\\
\bottomrule
\end{tabular}}
\end{table*}

\ul{\textbf{Developers show a lower likelihood of abandoning protestware.}}
Table \ref{tab:abandoned} offers insights into the characteristics of repositories influenced by protestware and the distribution of dependency changes.
We observe that the number of repositories dependent on the protestware, \caseOne, is greater than the vulnerability baseline, \caseThree.
For example, the number of dependent repositories for \caseOne~is 146, while the number for \caseThree~is 18.
Additionally, compared to the \caseThree, we find that the repositories affected by the protestware are relatively popular. 
The median values for the number of commits, stars, and contributors are 604, 124, and 23, respectively.

Table \ref{tab:abandoned} also reveals that on one hand, the majority of influenced developers do not abandon the protestware dependency. 
In \caseOne, only 15\% of repositories abandoned the protestware, which is similar to the \caseThree~and \caseFour~where only 6\% of repositories abandoned the vulnerable package.
There are several causes for why they were not abandoned. 
Different to a vulnerability, protestware has two key qualities:
\begin{itemize}
    \item The protester will not release a new version. Alternative mitigation include a new or forked library.
    \item The protestware is deemed benign and does not affect the library user. 
\end{itemize}
There also maybe other reasons that are unknown to the authors. 
For example, in the case of the \caseTwo, the manifest does not impact the behaviour of the software.
However, in the case of \caseOne, the reasons could be either the software is inactive or there is no alternative for mitigation, leaving the software hopeless to the remedy.
That being said, these are all points of discussion and we can only report the current situation. Probably this might be the reasoning why developers are less likely to make changes to the protestware package. 
For \caseThree, 82\% of the repositories and \caseFour, 54\% of the repositories addressed the vulnerability by upgrading to a newer version of the library.\\

\begin{tcolorbox}[colback=gray!5,colframe=awesome,title= RQ3 Summary]
Our results indicate that the affected repositories were relatively less likely to abandon the protestware when compared to vulnerabilities. For example, out of the 146 repositories that depend on \caseOne, 15\% have abandoned the dependency, which is higher than the 6\% abandonment rate for \caseThree~and \caseFour. This suggests that although there is some abandonment, it is not as significant as for vulnerabilities, possibly due to the lack of a defined mitigation strategy for protestware or because some protestware is deemed benign.
\end{tcolorbox}

% ======================================

\section{Lessons Learned}
Based on our results, we now discuss the implications and challenges for both developers and researchers.

\ul{\textbf{R1: Protestware discussions are multi-faceted and serve a diverse community of related entities}}. 

As demonstrated in our thematic analysis in RQ2b, it is evident that arguments from both sides are valid within their respective contexts (e.g., OSS philosophy and legal issues and rights). A solution would require a more comprehensive understanding of the bigger picture. Therefore, we urge the open-source community and all related stakeholders, including GitHub and open-source developers, to consider all viewpoints. These considerations should also be acknowledged by potential protesters and dogmatic developers.

Our taxonomy could be utilized to understand the possible narratives that the protestware could navigate and how they could be resolved. 
For example, most of the posts discuss rights and ethics, which could be discussed in a more open-minded way by the open-source community, especially understanding opposing thoughts.
One example is researchers and practitioners need to determine the most effective communication channels for developers to express their views to their OSS community.

\ul{\textbf{R2: Benign protestware is not abandoned}}. 
As demonstrated in RQ1, protestware is less likely to be abandoned. 
For example, in \caseOne, only 15\% of repositories abandoned the protestware.
Our results also indicate that protestware stimulates diverse discussions. These findings suggest that, when properly managed, protestware could be a viable approach to raising awareness of social needs, with developers becoming more receptive to individuals who want to highlight such issues.
However, it is still too early to draw definitive conclusions. 

In our case study, we find that the protester created a manifest file to document his protest.  
One idea is for GitHub to create a channel similar to the security.md policies file\footnote{\url{https://docs.github.com/en/code-security/getting-started/adding-a-security-policy-to-your-repository}}, so that protesters can peacefully protest in benign manner. 

\ul{\textbf{R3: Potential for Early Protestware Detection}}.
As shown in RQ1 (Table~\ref{tab:engagement}), we found that communication outlets and media platforms, such as Reddit, were able to quickly detect the protestware. Studies have demonstrated the effectiveness of NLP and sentiment analysis in detecting and analyzing social media content \citet{pak2010twitter,sarker2020cybersecurity}. One implication for researchers and practitioners is the need for tool support to monitor and promptly detect such disruptions before they spread throughout the ecosystem. Early detection can help users to prepare for the disruption, also the OSS community can prepare to handle the different conversations and narratives that might come into play during the protestware.

\ul{\textbf{R4: Protestware discussions contains technical mitigation instructions}}. 
Results from RQ2c indicate that many of the discussions likely include instructions (e.g., dependency management and alternative suggestions) on how to mitigate and resolve issues influenced by protestware.
An actionable implication could be that tool support is proposed to assist developers in migrating away from the protestware by utilizing the crucial information distilled from the discussions.
Similar to the vulnerability advisory, there could be a similar kind of board that enables developers to find mitigation strategies. 
Especially, using our taxonomy, such automatic tools similar to the Dependabot could be used to assist developers in quickly updating their dependencies or licenses.

\ul{\textbf{R5: Toxic conditions are a side-effect of protestware}}.
Manual analysis in RQ2a revealed that many of the opposing stances expressed by developers were toxic in nature. 
Several studies indicate that toxic interactions and uncivil language often lead to the failure of many new and longstanding OSS projects~\citep{miller2022did}.
Thus, to promote inclusivity and diversity in software development, we suggest implementing mechanisms or policies to prevent protestware from negatively impacting the OSS community. 
Recently, there has been a rise in tools like GitHub action bots used to automatically detect toxic comments in different OSS communities. 
Including the corpus of our protestware discussions would enhance the dataset used to train automatic toxicity tools.

\ul{\textbf{R6: Enhancing Security and Trust through Awareness}}. As discussed in RQ2b, our study underscores the importance of being aware of protestware and its implications. Specifically, 21\% of posts disseminate external links related to protestware, emphasizing the community's role in spreading awareness and information. Therefore, it is crucial to understand the implications of protestware. Practitioners can develop security protocols and training programs to better prepare their teams for potential protestware incidents, thereby enhancing overall software security and trust.

\ul{\textbf{R7: Promoting Sustainable Software Development Practices}}.
As highlighted in RQ2c, our study emphasizes the need for sustainable software development practices that can withstand influences like protestware. The results in Table \ref{tab:abandoned} show that a significant percentage of developers do not abandon protestware, with only 15\% of repositories abandoning \caseOne, suggesting the necessity for sustainable practices to handle such scenarios, considering that some of these repositories may be inactive. Practitioners should focus on creating robust and flexible development environments that can adapt to changes and challenges. By implementing sustainable practices, organizations can ensure the long-term health and stability of their software projects, even in the face of unexpected disruptions.

\section{Threats to Validity }
\label{sec:threats}
One of {\textit{external threat}} to the validity of our findings is the ability to generalize them, as our work is presented as a multiple case study, encompassing \caseOne~and \caseTwo.
We acknowledge that observations may not be generalized to other protestware cases and ecosystems.
As the first study to systematically analyze the discussions of protestware, we argue that these two cases can be pioneering baselines that represent society issues of economic and world politics. We envision more studies to follow our work, exploring how to reproduce our results on other protestware in the future. 

%  Another \textit{external threat} to the validity of our findings is the risk of false positives in our quantitative results for RQ1, as it may lead to overestimation. We acknowledge this risk but believe it was the most accurate measurement of the spread of protestware discussion through intentional linking and mentioning in related posts across GitHub.
In terms of {\textit{construct threats}}, a major threat arises on how we systematically retrieve posts from both GitHub and social media.
We understand that a different methodology, such as different search methods may change the amount of data collected.
Furthermore, we only consider two platforms: Reddit and Hacker News, and it is possible that we may have an underestimation of the true impact.
Although twitter (X) is more prevalent, due to recent events, we were unable to utilize X new API to collect information.\footnote{\url{https://developer.twitter.com/en/docs/twitter-api/migrate/whats-new}}
That being said, we selected a method that would return fewer false positives. 
Hence, as future work, we plan to expand our work to include a more robust and dynamic method to collect our dataset. 
Additionally, a \textit{construct threat} in our study arises from the potential subjectivity and bias introduced during the coding process. Despite our collaborative efforts to develop and refine the code schema, the inherent subjectivity in qualitative coding means that independent coders might find it challenging to replicate our taxonomy and achieve consistent results. This poses a limitation to the replicability of our findings.

We identify two key {\textit{internal threats}}.
The first threat exists in the selection of the metrics to measure the disruption of protestware in RQ1.
Although the engagement metric quantifies speed, we cannot make any statistical claims.
The second threat occurs during our multiple manual analyses of posts in RQ2, where the codes or themes may be mislabeled due to their subjective nature.
While we acknowledge that qualitative measures are prone to subjective bias, we implemented a typical systematic method established in prior works and involved more than one coder in each manual analysis.

\section{Related work}
We now position our work with respect to the related literature.

\paragraph{\textbf{Software Development Reliance on Software Ecosystems}} Modern software development heavily relies on third-party libraries, which is evident by supply chain attacks~\citep{vu2020towards, zahan2022weak, zimmermann2019small}.
A recent report from \citet{report} stated that supply chain attacks have dramatically increased 650\% in 2021 on top of year-over-year growth of 430\% in 2020.
\citet{ohm2020backstabber} studied the three popular package ecosystems (npm, PyPI, and RubyGems) and presented a taxonomy of attack vectors and a
dataset of malicious software packages. 
\citet{hou2023systematic} demonstrated that stakeholders' trust is frequently violated by bad actors and crippling vulnerabilities in the software supply chain.
At the same time, several automated solutions have been proposed to protect against supply chain attacks~\citep{ohm2020towards, scalco2022feasibility}.
There have also been several tools proposed~\citep{gonzalez2021anomalicious,sejfia2022practical} for automatically detecting malicious npm packages and their evaluation results show promise in practice.

\paragraph{\textbf{Economic Incentives in Open Source}}
Bounties are frequently used to attract developers and motivate them to complete diverse software development tasks, such as software vulnerabilities~\citep{finifter2013empirical} and bug reports~\citep{hata2017understanding}.
\citet{krishnamurthy2006bounty} provided an overview of bounty programs in FLOSS and suggested that bounty hunters' responses are related to the workload.
\citet{finifter2013empirical} analyzed vulnerability rewards programs for
Chrome and Firefox and reported that such programs are economically effective, compared to the cost of hiring full-time security researchers.
\citet{maillart2017given} recommended that project managers should dynamically adjust the value
of rewards according to the market situation.
\citet{zhou2020studying} studied the association between bounties and the issue-addressing likelihood of
GitHub issue reports.
% Their findings indicated that issue reports are more likely to be addressed if the bounties are used commonly and offered early.
In the context of the question-answering process, \citet{zhou2020bounties} observed that questions
are likely to attract more traffic after receiving a bounty.
The latest works~\citep{shimada2022github, zhou2022studying} also investigate GitHub Sponsors, a service allowing developers to accept donations.
Specifically, \citet{shimada2022github} found that sponsored developers are more active than non-sponsored developers.

%\subsection{Social Good, Toxicity, and Codes of Conduct in OSS}
%\raula{please add a section here}
\paragraph{\textbf{Social Good in Software Engineering}}
Based on the positive impact of software on society in recent years, \citet{10.1145/2591062.2591121} defined SE for ``social good'' as the development, maintenance, and sustainability of software aimed at promoting social change.
Along with this concept, \citet{9402112} specifically targeted Open Source Software for Social Good (OSS4SG), which refers to OSS projects that benefit society.
In April 2020, GitHub's Social Impact Sector released a report indicating a tremendous increase in demand for open-source projects with social impact, and a lack of research and resources related to them. 
\citet{dekhtyar2020re} recommend that SE conferences should focus on activities that benefit society and that these activities should be an integral, valued, and recognized part of the conference programs.
Protestware provides a different lens by which we can view different relationships of OSS to such societal issues. 

\section{Conclusion and Future Work}
At this early stage, it is unsure whether or not protestware is here to stay in the open source software ecosystem. 
With the heavy reliance on the ecosystem and its supply chain, the mere presence of protestware is a reminder that software development is becoming more and more intertwined with society and its issues, ranging from economic to political. 
In this first large-scale exploration into protestware, we present the different narratives that drive this emerging phenomenon. 
The results demonstrate that developers are less likely to abandon the protestware.
Developers are often vocal about their opinions on protestware and opinions are multifaceted with different contexts, perceptions, and tones. 
Additionally, they frequently provide technical information that can include mitigation instructions.

This work lays the foundation for further research and provides various opportunities. 
These future research opportunities encompass not only providing developers with insights into the varying perspectives but also offering actionable guidance to foster constructive dialogue and reduce toxicity in these sensitive discussions. 
Such frameworks could transform protestware from a potential point of contention into an instrument for positive social change, aligning the open-source community with broader societal values and goals.

\section*{Acknowledgement}
This work was supported by JST SICORP Grant Number JPMJSC2206, JST BOOST Grant Number JPMJBS2423, and Japanese Society for the Promotion of Science (JSPS) KAKENHI grants (JP23K16864), and

\section*{Data Availability Statements}

The datasets generated during and/or analysed during the current study are available at \DOIbox{10.5281/zenodo.8207976}

\section*{Declarations}

\subsection*{Conflict of Interest}
The authors declare that Raula Gaikovina Kula, Hideaki Hata, and Christoph Treude are members of the EMSE Editorial Board. 
All co-authors have seen and agree with the contents of the manuscript and there is no financial interest to report.

\bibliographystyle{spbasic}      % basic style, author-year citations {ieeetr}
\typeout{}
\bibliography{bibliography}   % name your BibTeX data base

\vspace{3\baselineskip}

{\setlength\intextsep{0pt}
\begin{wrapfigure}{l}{25mm} 
    \includegraphics[width=1in,height=1.25in,clip,keepaspectratio]{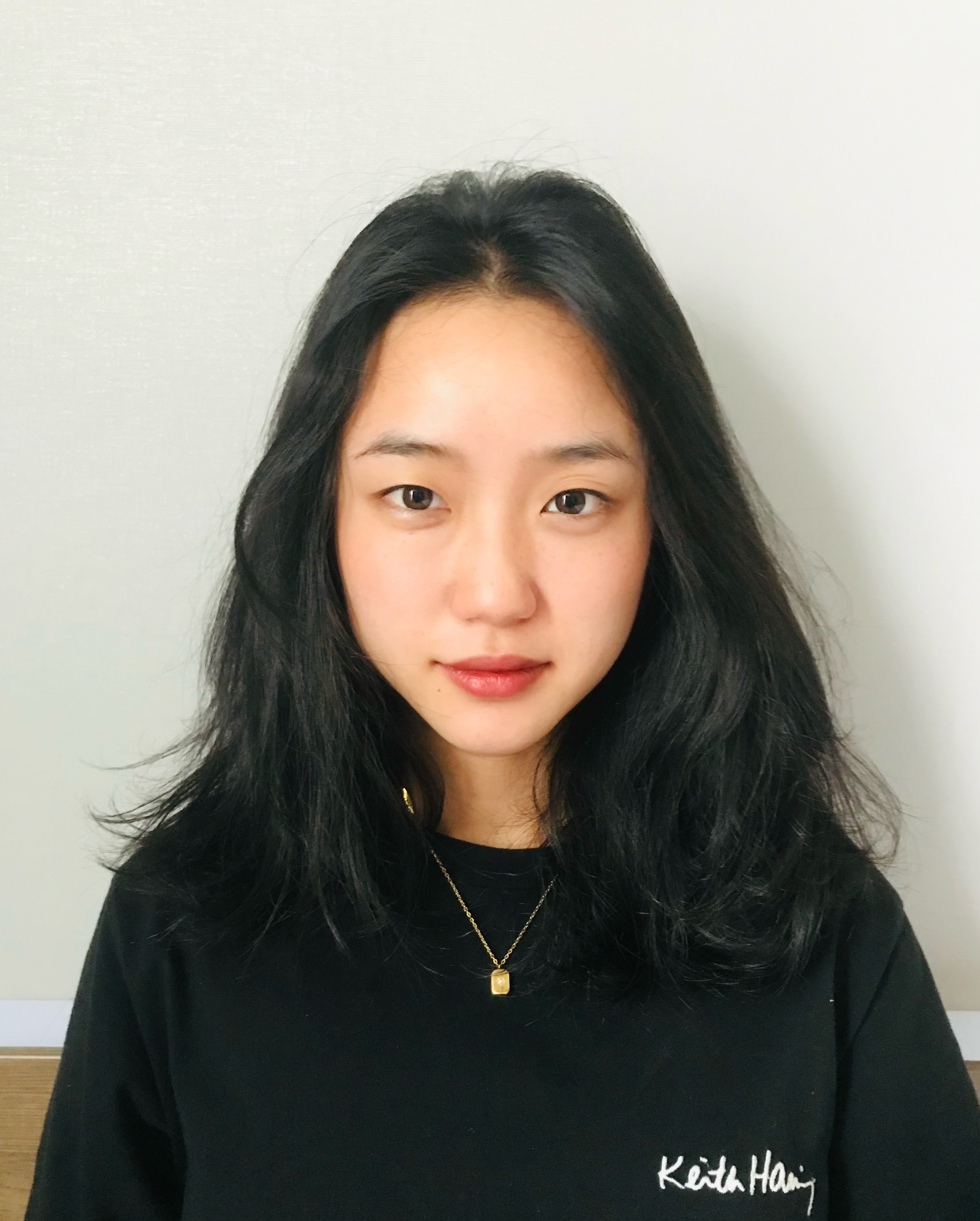}
\end{wrapfigure}\par
\noindent\textbf{Youmei Fan}\\ is a Ph.D. student in Software Engineering Laboratory at the Graduate School of Science and Technology, Nara Institute of Science and Technology (NAIST). She completed a master's course in Software Engineering Laboratory at the Graduate School of Science and Technology, Nara Institute of Science and Technology (NAIST), Japan. During her master's degree, her main research interests related to the human aspect of software engineering, open-source software sustainability, and data mining.
Find her at \url{https://www.linkedin.com/in/youmei-fan-513a161a6/}.
\par}

\vspace{2\baselineskip}

{\setlength\intextsep{0pt}
\begin{wrapfigure}{l}{25mm} 
    \includegraphics[width=1in,height=1.25in,clip,keepaspectratio]{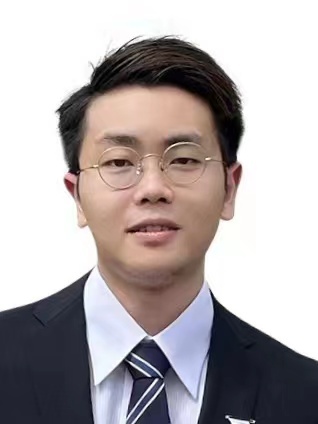}
\end{wrapfigure}\par
\noindent\textbf{Dong Wang}\\ is currently an associate professor at Tianjin University, China.
Prior to this, he worked as an assistant processor at Kyushu University, Japan.
He received Ph.D. degree from Nara Institute of Science and Technology, Japan. He is a member of the IEEE. His research interests include mining software repositories, empirical software engineering, human aspects, and AI4SE. More about his information is available online at \url{https://dong-w.github.io/}. \par}

\vspace{2\baselineskip}

{\setlength\intextsep{0pt}
\begin{wrapfigure}{l}{25mm} 
    \includegraphics[width=1in,height=1.25in,clip,keepaspectratio]{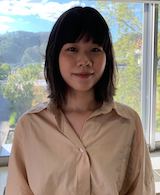}
\end{wrapfigure}\par
\noindent\textbf{Supatsara Wattanakriengkrai}\\ is a Ph.D. student in Software Engineering Laboratory at the Graduate School of Science and Technology, Nara Institute of Science and Technology (NAIST), Japan. She received a Master's degree from NAIST in 2021. Her main research interests include empirical software engineering, mining software repositories, and library dependencies in software ecosystems.
More about her work is available online at \url{https://www.supatsarawat.com/}. \par}

\vspace{2\baselineskip}

{\setlength\intextsep{0pt}
\begin{wrapfigure}{l}{25mm} 
    \includegraphics[width=1in,height=1.25in,clip,keepaspectratio]{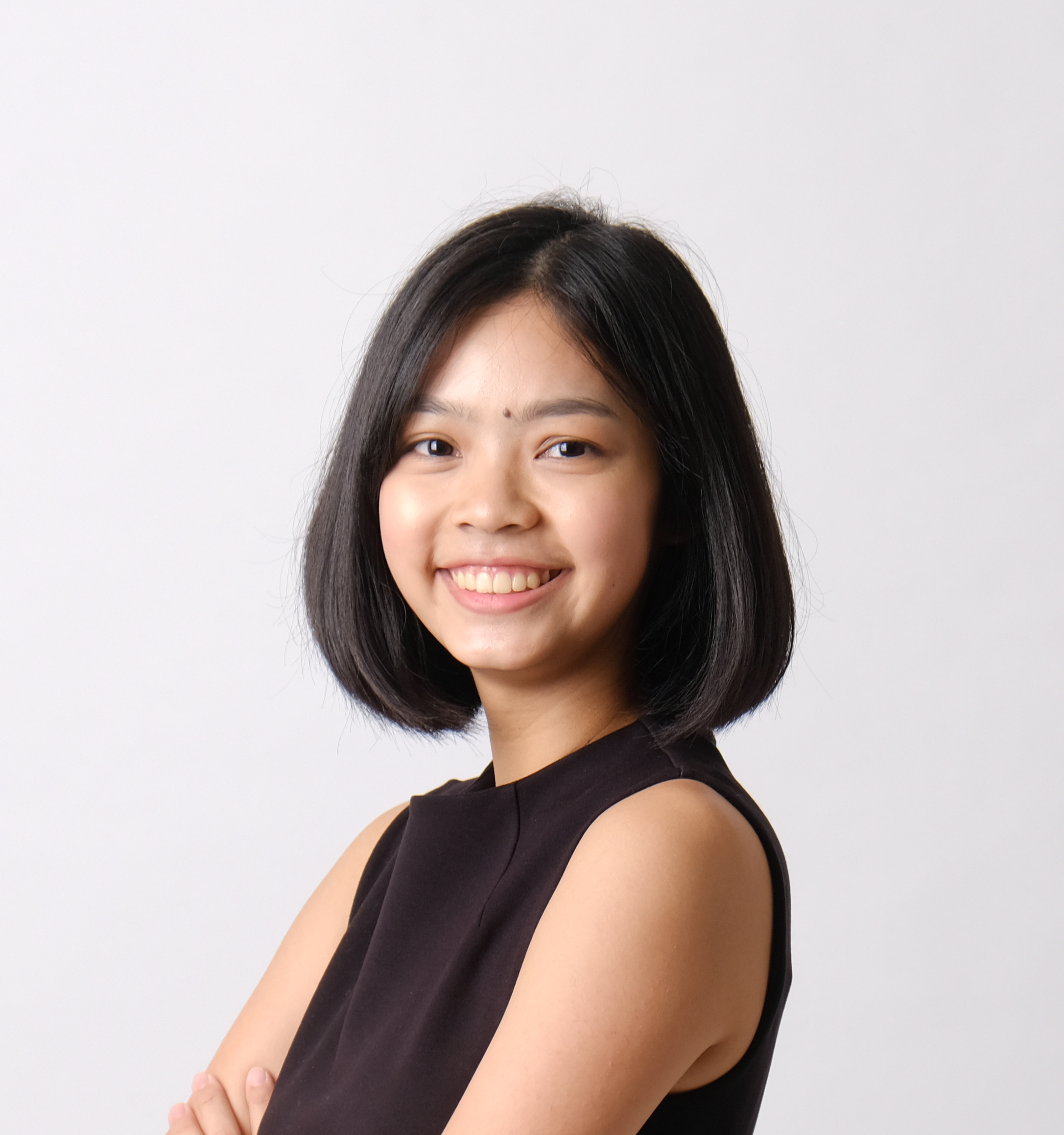}
\end{wrapfigure}\par
\noindent\textbf{Hathaichanok Damrongsiri}\\ is a Master's student in the Software Engineering Laboratory at the Graduate School of Science and Technology, Nara Institute of Science and Technology (NAIST), Japan. She received a bachelor's degree in Software Engineering from Chiang Mai University and remains interested in the field of Software Engineering, especially in empirical software engineering, software educational materials, and teaching methods. Find her at \url{https://www.linkedin.com/in/ploychanok/} \par}

\vspace{2\baselineskip}

{\setlength\intextsep{0pt}
\begin{wrapfigure}{l}{25mm} 
    \includegraphics[width=1in,height=1.25in,clip,keepaspectratio]{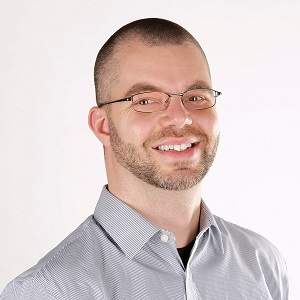}
\end{wrapfigure}\par
\noindent\textbf{Christoph Treude}\\ is a Senior Lecturer in Software Engineering in the School of
Computing and Information Systems at the University of Melbourne. The goal
of his research is to improve the quality of software and the productivity of
those producing it, with a particular focus on getting information to software
developers when and where they need it. He has authored more than 100
scientific articles with more than 200 co-authors, and his work has received
an ARC Discovery Early Career Research Award (2018-2020), industry funding
from Google, Facebook, and DST, as well as four best paper awards including
two ACM SIGSOFT Distinguished Paper Awards. He currently serves as a board
member on the Editorial Board of the Empirical Software Engineering journal
and was general co-chair for the 36th IEEE International Conference on Software
Maintenance and Evolution. \par}

\vspace{2\baselineskip}

{\setlength\intextsep{0pt}
\begin{wrapfigure}{l}{25mm} 
    \includegraphics[width=1in,height=1.25in,clip,keepaspectratio]{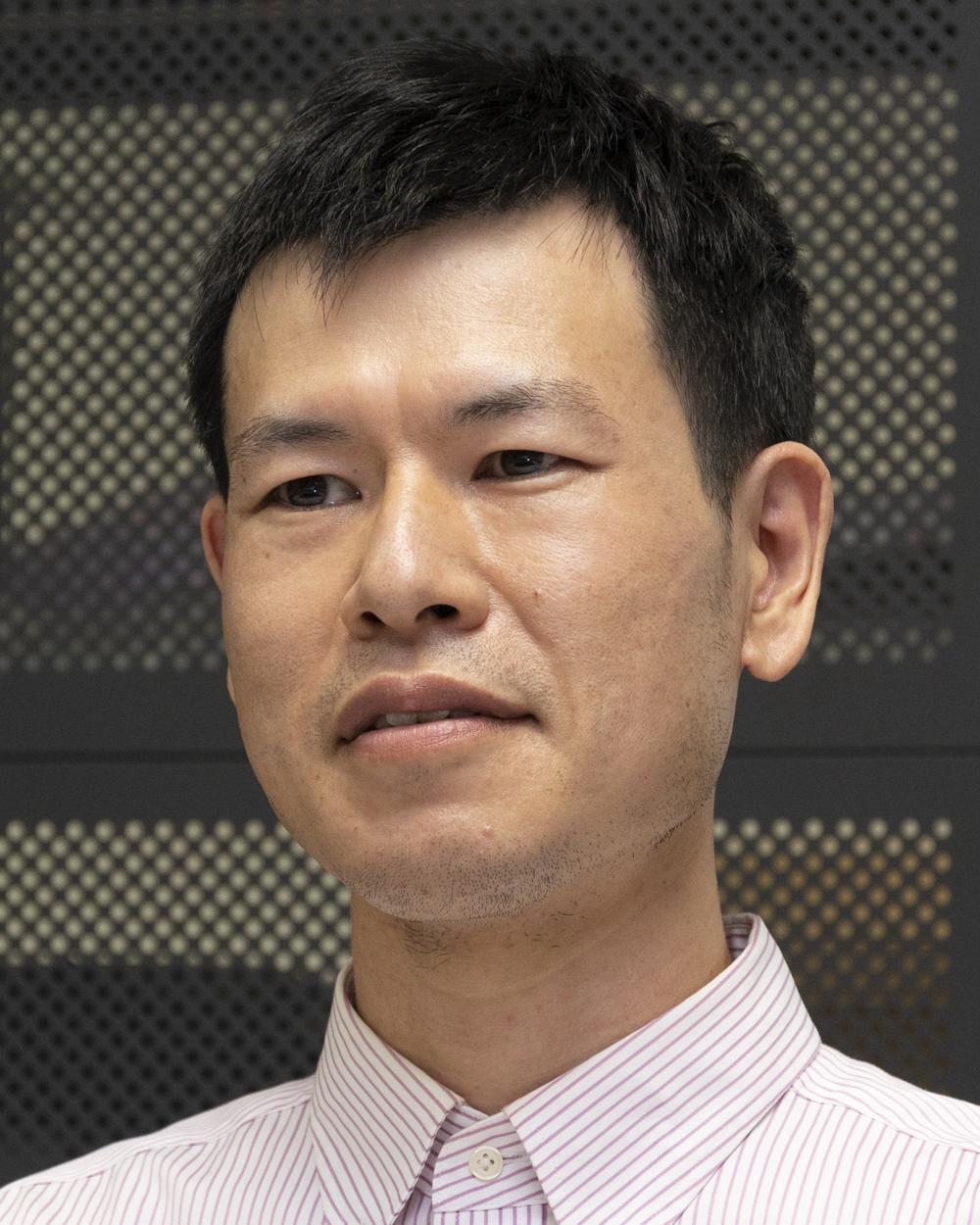}
\end{wrapfigure}\par
\noindent\textbf{Hideaki Hata}\\ is an Associate Professor at Shinshu University. He received his Ph.D. in information science from Osaka University. His research interests include software ecosystems, human capital in software engineering, and software economics. More about Hideaki and his work is available online at \url{https://hideakihata.github.io/}. \par}

\vspace{2\baselineskip}

{\setlength\intextsep{0pt}
\begin{wrapfigure}{l}{25mm} 
    \includegraphics[width=1in,height=1.25in,clip,keepaspectratio]{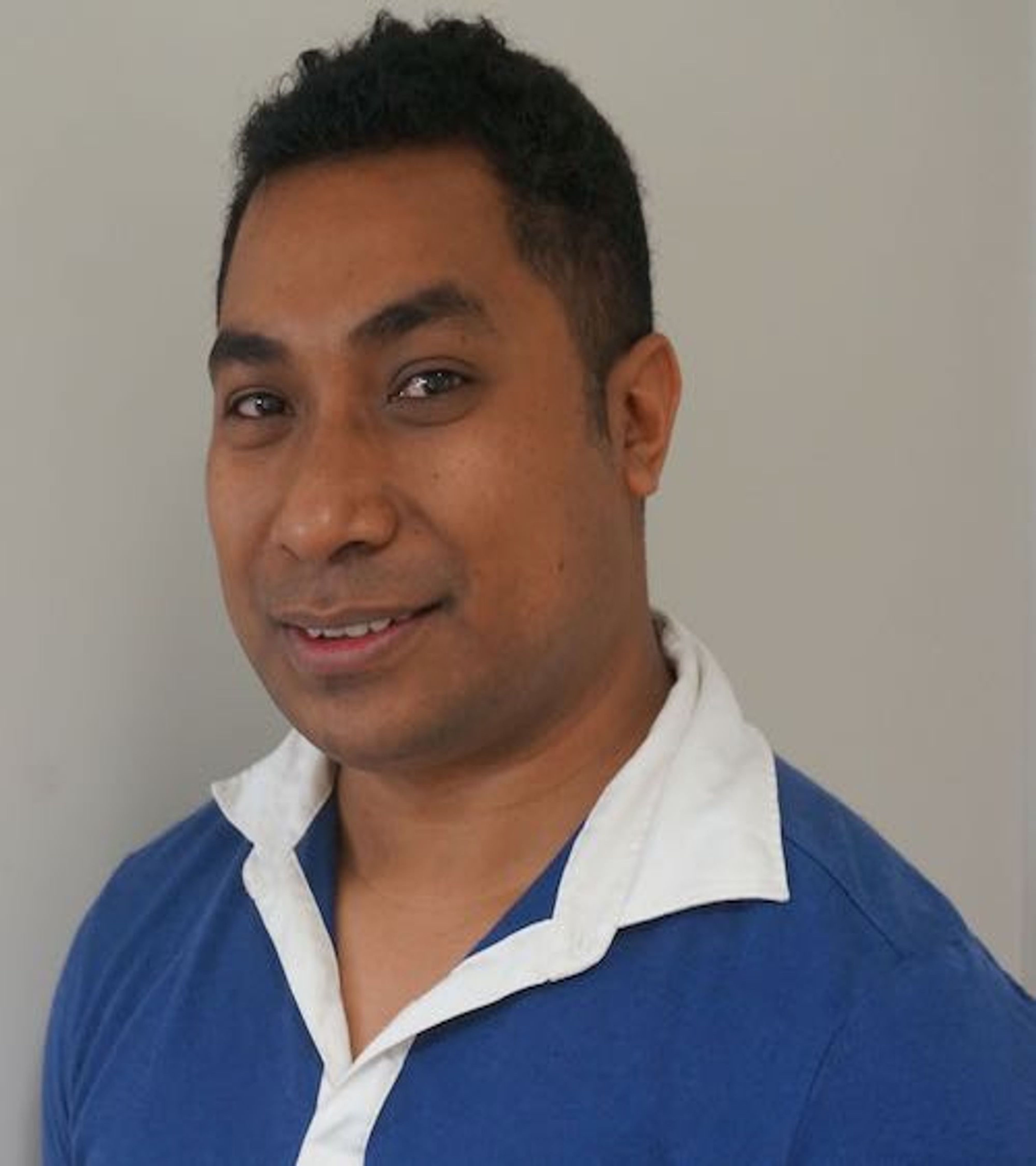}
\end{wrapfigure}\par
\noindent\textbf{Raula Gaikovina Kula}\\ is an associate professor at the Nara Institute of Science
and Technology (NAIST), Japan. He received his Ph.D. degree from NAIST in 2013
and was a Research Assistant Professor at Osaka University. He is active in the
Software Engineering community, serving the community as a PC member for
premium SE venues, some as organising committee, and reviewer for journals.
His current research interests include library dependencies and security in the
software ecosystem, program analysis such as code clones, and human aspects such as code reviews and coding proficiency. Find him at \url{https://raux.github.io/}
and @augaiko on Twitter. Contact him at raula-k@is.naist.jp. \par}

\end{document}